\documentclass[12pt,a4paper]{article}
\usepackage{amssymb,amsmath}
\usepackage{color} 
\usepackage{graphicx}

\newcommand{\bc}{\begin{center}}
\newcommand{\ec}{\end{center}}
\newcommand{\bd}{\begin{displaymath}}
\newcommand{\ed}{\end{displaymath}}
\newcommand{\be}{\begin{equation}}
\newcommand{\ee}{\end{equation}}
\newcommand{\ba}{\begin{array}}
\newcommand{\ea}{\end{array}}
\newcommand{\bt}{\begin{tabular}}
\newcommand{\et}{\end{tabular}}

\begin{document}

\begin{titlepage}


\begin{center}
{
\sffamily
\LARGE
Leptogenesis and Dark Matter-Nucleon Scattering\\[2mm]
 Cross Section in the SE$_6$SSM}\\[8mm]
{\large Roman Nevzorov\\[3mm]
\itshape{I. E. Tamm Department of Theoretical Physics,}\\[0mm]
\itshape{Lebedev Physical Institute, Leninsky prospect 53, 119991 Moscow, Russia}
}\\[1mm]
\end{center}
\vspace*{0.75cm}

\begin{abstract}{
\noindent
The $E_6$ inspired extension of the minimal supersymmetric (SUSY) standard model (MSSM) with an extra $U(1)_{N}$ gauge symmetry, under
which right-handed neutrinos have zero charge, involves exotic matter beyond the MSSM to ensure anomaly cancellation. We consider the variant of
this extension (SE$_6$SSM) in which the cold dark matter is composed of the lightest neutral exotic fermion and gravitino. The observed baryon
asymmetry can be induced in this case via the decays of the lightest right--handed neutrino/sneutrino into exotic states even for relatively
low reheating temperatures $T_R\lesssim 10^{6-7}\,\mbox{GeV}$. We argue that there are some regions of the SE$_6$SSM parameter space, which are safe
from all current constraints, and discuss the implications of this model for collider phenomenology.
}
\end{abstract}

\end{titlepage}

\newpage
\section{Introduction}

The observed baryon asymmetry and the presence of cold dark matter in the Universe stimulates the investigation
of extensions of the Standard Model (SM). New physics beyond the SM permits to induce the baryon asymmetry if
Sakharov conditions are fulfilled \cite{Sakharov:1967dj}. The proposed new physics scenarios include baryogenesis in
Grand Unified theories (GUTs) \cite{gut1}--\cite{gut7}, the Affleck-Dine mechanism \cite{AD1,AD2},
baryogenesis via leptogenesis \cite{Fukugita:1986hr}, electroweak baryogenesis \cite{ew-baryogen}--\cite{Huber:2000mg}, etc\,.
In the case of thermal leptogenesis \cite{Fukugita:1986hr} lepton asymmetry is generated due to the decays of the lightest
right--handed neutrino. The realisation of this mechanism within the type I seesaw models \cite{Minkowski:1977sc}, in which
CP and lepton number are violated, allows for understanding of the mass hierarchy in the lepton sector
if the right--handed neutrinos are superheavy. In this scenario the induced lepton asymmetry gets partially converted into
baryon asymmetry via sphaleron processes \cite{Kuzmin:1985mm,Rubakov:1996}.

After inflation in the reheating epoch, which is characterized by a reheat temperature $T_R$, the right--handed neutrinos can be
produced by thermal scattering if $T_R > M_1$. In the SM and minimal supersymmetric (SUSY) standard model (MSSM) such production
process results in the appropriate baryon asymmetry only when the lightest right--handed neutrino mass $M_1$ is larger than
$10^9\,\mbox{GeV}$ \cite{lb1,lb2}. Therefore thermal leptogenesis in the MSSM and its extensions may take place when
$T_R\gtrsim 10^9\,\mbox{GeV}$. This lower bound on the reheat temperature leads to the gravitino problem \cite{grav-prob1,grav-prob2}
in the supergravity (SUGRA) models, that lead to the sparticle mass scale below $10\,\mbox{TeV}$.
Indeed, so high $T_R$ gives rise to an overproduction of gravitinos. Since gravitinos are sufficiently longlived they
tend to decay after Big Bang Nucleosynthesis (BBN). Such decays destroy the agreement between the predicted and observed light
element abundances. To preserve the success of BBN the relic abundance of gravitinos has to be relatively small.
It becomes low enough if reheat temperature is lower than $10^{6-7}\,\mbox{GeV}$ \cite{tr1}--\cite{tr2}.

In this context it seems to be interesting to study the generation of matter–-antimatter asymmetry and formation of cold dark matter
in the framework of well motivated $E_6$ inspired extensions of the SM. In the $E_6$ inspired composite Higgs model (E$_6$CHM) \cite{Nevzorov:2015sha,Nevzorov:2016fxp} the process of the baryon asymmetry generation was explored
in \cite{Nevzorov:2017rtf,Nevzorov:2022zjo}. The $E_6$ inspired $U(1)$ extensions of the MSSM
implies that near the GUT scale $M_X$ the $E_6$ gauge group is broken down to $SU(3)_C\times SU(2)_W\times U(1)_Y\times U(1)'$
(for review see \cite{Hewett:1988xc,Langacker:2008yv}) where $SU(3)_C\times SU(2)_W\times U(1)_Y$ is the SM gauge group and
\begin{equation}
U(1)'=U(1)_{\chi}\cos\theta_{E_6} + U(1)_{\psi}\sin\theta_{E_6}\,.
\label{1}
\end{equation}
In Equation (\ref{1}) the $U(1)_{\psi}$ and $U(1)_{\chi}$ symmetries are associated with the subgroups $E_6 \supset SO(10) \times U(1)_{\psi}$
and $SO(10) \supset SU(5) \times U(1)_{\chi}$. The $E_6$ inspired $U(1)$ extensions of the MSSM can originate from the heterotic superstring
theory with $E_8\times E'_8$-gauge symmetry. Some phenomenological consequences of the heterotic string model were considered in~\cite{Khlopov:2002gg,Khlopov:2021xnw}.

Within the SUSY models with extra $U(1)'$ the anomalies are canceled if the particle spectrum contains complete
representations of $E_6$. Because of this the particle spectrum in the models under consideration is usually extended by the supermultiplets
of exotics so that it consists of three 27-dimensional representations of $E_6$ ($27_i$ with $i=1,2,3$) at low energies. These
$27$--plets decompose under $SU(5)\times U(1)_{\psi}\times U(1)_{\chi}$ as follows:
\begin{equation}
\begin{array}{rcl}
27_i &\to &
\left(10,\,\dfrac{1}{\sqrt{24}},\,-\dfrac{1}{\sqrt{40}}\right)_i
+\left(5^{*},\,\dfrac{1}{\sqrt{24}},\,\dfrac{3}{\sqrt{40}}\right)_i
+\left(5^{*},\,-\dfrac{2}{\sqrt{24}},\,-\dfrac{2}{\sqrt{40}}\right)_i
\\[3mm] & + &
\left(5,\,-\dfrac{2}{\sqrt{24}},\,\dfrac{2}{\sqrt{40}}\right)_i
+\left(1,\,\dfrac{4}{\sqrt{24}},\,0\right)_i
+\left(1,\,\dfrac{1}{\sqrt{24}},\,-\dfrac{5}{\sqrt{40}}\right)_i\,.
\end{array}
\label{2}
\end{equation}
Here the first, second and third quantities in brackets are the $SU(5)$
representation as well as $U(1)_{\psi}$ and $U(1)_{\chi}$ charges.
The SM family, which consists of the doublets of left--handed quarks
$Q_i$ and leptons $L_i$, right-handed up-- and down--quarks ($u^c_i$ and
$d^c_i$) as well as right--handed charged leptons $(e^c_i)$, corresponds
to $\left(10,\,\dfrac{1}{\sqrt{24}},\,-\dfrac{1}{\sqrt{40}}\right)_i$ +
$\left(5^{*},\,\dfrac{1}{\sqrt{24}},\,\dfrac{3}{\sqrt{40}}\right)_i$.
The last term in Equation (\ref{2}), $\left(1,\,\dfrac{1}{\sqrt{24}},\,-\dfrac{5}{\sqrt{40}}\right)_i$,
represents the right-handed neutrinos $N^c_i$. The
next-to-last term, $\left(1,\,\dfrac{4}{\sqrt{24}},\,0\right)_i$,
is associated with new SM-singlet fields $S_i$, that carry non-zero $U(1)_{\psi}$
charges. The $SU(2)_W$--doublets ($H^d_{i}$ and $H^u_{i}$) from
$\left(5^{*},\,-\dfrac{2}{\sqrt{24}},\,-\dfrac{2}{\sqrt{40}}\right)_i$
and $\left(5,\,-\dfrac{2}{\sqrt{24}},\,\dfrac{2}{\sqrt{40}}\right)_i$
have the quantum numbers of the MSSM Higgs doublets. The colour triplets from
these $SU(5)$ supermultiplets are associated with exotic quarks $\overline{D}_i$ and $D_i$
with electric charges $+ 1/3$ and $-1/3$ respectively. They carry
a $B-L$ charge $\left(\pm\dfrac{2}{3}\right)$ which is twice larger than
the $B-L$ charges of ordinary quarks.

Among the $E_6$ inspired $U(1)$ extensions of the MSSM there is a unique combination of $U(1)_{\psi}$ and $U(1)_{\chi}$
corresponding to $\theta_{E_6}=\arctan\sqrt{15}$ for which $N^c_i$ do not participate in the gauge interactions.
Only in this SUSY model with extra $U(1)_{N}$ gauge symmetry, i.e. the so-called Exceptional Supersymmetric Standard Model (E$_6$SSM)
\cite{King:2005jy,King:2005my} (for recent review see \cite{e6ssm-sym}), the right--handed neutrinos can be rather heavy
inducing the mass hierarchy in the lepton sector. The heavy Majorana right--handed neutrinos are allowed to decay into final states
with lepton number $L=\pm 1$, resulting in lepton and baryon asymmetries in the early Universe
\cite{Hambye:2000bn}--\cite{Nevzorov:2018leq}. In the $U(1)_{N}$ extensions of the MSSM the extra exotic states with the TeV scale masses
can give rise to rapid proton decay and flavor-changing transitions. The corresponding operators can be suppressed in the E$_6$SSM using
a set of discrete symmetries \cite{King:2005jy,King:2005my}.

In this article we focus on the variant of the E$_6$SSM (SE$_6$SSM) in which a single $\tilde{Z}^{H}_2$ symmetry forbids non-diagonal
flavor transitions and most dangerous operators that violate baryon and lepton numbers. In the next section the SE$_6$SSM is specified.
In section 3 the thermal leptogenesis within this $U(1)_{N}$ extension of the MSSM is considered. The interactions of the dark matter
states with the nucleons is explored in section 4. In Section 5 we summarize the results of our studies and discuss the implications of
the SUSY model under consideration for collider phenomenology.

\section{The $U(1)_N$ extension of the MSSM with exact custodial $\tilde{Z}^{H}_2$ symmetry}

At very high energies the $E_6$ orbifold SUSY GUTs can be reduced to an effective rank--6 SUSY model
based on the $SU(3)_{C}\times SU(2)_{W}\times U(1)_{Y}\times U(1)_{\chi}\times U(1)_{\psi}$ gauge symmetry \cite{Nevzorov:2012hs}.
If the particle content of this rank--6 model involves just three $27$--plets at low energies then the most general renormalisable
superpotential comes from the $27\times 27\times 27$ decomposition of $E_6$ and can be written as
\begin{equation}
\begin{array}{rcl}
W_{E_6} & = & W_0+W_1+W_2\,,\\
W_0 & = & \lambda_{ijk} S_i (H^d_{j} H^u_{k})+\kappa_{ijk} S_i (D_j \overline{D}_k)+
h^N_{ijk} N_i^c (H^u_{j} L_k)+ h^U_{ijk} u^c_{i} (H^u_{j} Q_k) +\\[0mm]
& + & h^D_{ijk} d^c_i (H^d_{j} Q_k) +  h^E_{ijk} e^c_{i} (H^d_{j} L_k)\,,\\[0mm]
W_1&=& g^Q_{ijk} D_{i} (Q_j Q_k)+g^{q}_{ijk}\overline{D}_i d^c_j u^c_k\,,\\[0mm]
W_2&=& g^N_{ijk}N_i^c D_j d^c_k+g^E_{ijk} e^c_i
D_j u^c_k+g^D_{ijk} (Q_i L_j) \overline{D}_k\,,
\end{array}
\label{3}
\end{equation}
where the summation over repeated indexes is implied and $i,j,k=1,2,3$.

From Equation (\ref{3}) it follows that if all Yukawa couplings in $W_1$ and $W_2$ have non--zero values
then one cannot define the baryon ($B$) and lepton ($L$) numbers so that the Lagrangian of this model
is invariant under the corresponding $U(1)_{B}$ and $U(1)_{L}$ global symmetries.
Therefore, as in the simplest SUSY extensions of the SM, the gauge symmetry in the $E_6$ inspired SUSY models
does not forbid the operators which violate lepton and baryon numbers. This means that in general
these models lead to rapid proton decay. Moreover, since three pairs of $H^u_{i}$ and $H^d_{i}$ couple to
charged leptons and ordinary quarks the corresponding Yukawa interactions may give rise to unacceptably
large flavor changing processes at the tree level. In particular, these interactions can induce
new channels of muon decay such as $\mu\to e^{-}e^{+}e^{-}$ and contribute to the amplitude of
$K^0$--$\overline{K}^0$ oscillations.

Although $U(1)_{B}$ and $U(1)_{L}$ symmetries are not conserved the superpotential (\ref{3})
possesses $U(1)_{B-L}$ symmetry associated with $B-L$ number conservation if the exotic quark supermultiplets
$\overline{D}_i$ ($D_i$) carry $B-L$ numbers $B-L=\dfrac{2}{3}\Biggl(-\dfrac{2}{3}\Biggr)$. As a consequence
$U(1)_{\chi}\times U(1)_{\psi}$ gauge symmetry can be broken down to matter parity $Z_{2}^{M}=(-1)^{3(B-L)}$,
which is a discrete subgroup of $U(1)_{B-L}$. In the case of the E$_6$SSM $U(1)_{\chi}\times U(1)_{\psi}$
symmetry is expected to be broken to $U(1)_{N}\times Z_{2}^{M}$ near the GUT scale $M_X$ \cite{King:2005jy,King:2005my}.
Such breakdown can be attained if $N^c_H$ and $\overline{N}_H^c$ components of some extra $27_H$ and $\overline{27}_H$
representations develop vacuum expectation values (VEVs) along the $D$-flat direction $<N_H^c>=<\overline{N}_H^c>$ \cite{Nevzorov:2012hs}.
These VEVs may also induce the Majorana mass terms of the right--handed neutrinos (i.e. $\dfrac{1}{2}M_{ij} N_i^c N_j^c$) in the
superpotential with the intermediate scale mass parameters $M_{ij}$ through the non-renormalizable operators of the form
\begin{equation}
\delta W=\dfrac{\kappa_{ij}}{M_{Pl}}(\overline{27}_H\, 27_i)(\overline{27}_H\, 27_j)\qquad \Longrightarrow\qquad
M_{ij}=\dfrac{2\kappa_{ij}}{M_{Pl}}<\overline{N}_H^c>^2\,,
\label{4}
\end{equation}
where $M_{Pl}=(8\pi G_N)^{-1/2} \simeq 2.4\cdot 10^{18}\,\mbox{GeV}$~ is~ the~ reduced~ Planck~ mass.~
When $<N_H^c>\simeq <\overline{N}_H^c> \simeq M_X\simeq 2-3\cdot 10^{16}\,\mbox{GeV}$, the observed pattern of masses
and mixing angles of the left--handed neutrinos can be obtained.

Over the last fifteen years, several modifications of the E$_6$SSM, in which the operators leading to
rapid proton decay and flavor changing processes are suppressed, have been proposed
\cite{King:2005jy},\cite{King:2005my},\cite{Nevzorov:2012hs}--\cite{Nevzorov:2022zns}.
The implications of the $U(1)_{N}$ extensions of the MSSM were explored for $Z'$ physics~\cite{Suematsu:1997au},
neutralino sector \cite{Suematsu:1997au}--\cite{Keith:1996fv},
electroweak (EW) symmetry breaking (EWSB) \cite{Keith:1997zb},\cite{Suematsu:1994qm}--\cite{Daikoku:2000ep},
the renormalization group (RG) flow of couplings~\cite{Keith:1997zb,King:2007uj},
the renormalization of VEVs~\cite{Sperling:2013eva,Sperling:2013xqa},
non-standard neutrino models~\cite{Ma:1995xk} and
dark matter~\cite{Khalil:2020syr}--\cite{Nevzorov:2022zns},\cite{Hall:2009aj}--\cite{Athron:2016qqb}.
Within the E$_6$SSM the upper bound on the lightest Higgs mass near the quasi-fixed point was studied
in~\cite{Nevzorov:2013ixa}. This quasi-fixed point is an intersection of
the invariant and quasi--fixed lines~\cite{Nevzorov:2001vj,Nevzorov:2002ub}.
The particle spectrum and corresponding phenomenological implications in the constrained E$_6$SSM
(cE$_6$SSM) and its modifications were analyzed in \cite{Athron:2015vxg}--\cite{Athron:2012sq}.
The degree of fine tuning and threshold corrections were examined in \cite{Athron:2013ipa,Athron:2015tsa}
and \cite{Athron:2012pw} respectively. In the E$_6$SSM extra exotic matter may lead to distinctive LHC signatures
~\cite{King:2005jy}--\cite{King:2005my},\cite{Howl:2007zi},\cite{Athron:2010zz},\cite{King:2006vu}--\cite{Belyaev:2012jz}
and can give rise to non--standard decays of the lightest Higgs boson
~\cite{Athron:2014pua},\cite{Hall:2010ix},\cite{Nevzorov:2013tta}--\cite{Nevzorov:2020jdq}.

In addition to three complete $27$--plets the splitting of bulk $27^{\prime}$ supermultiplets in the $E_6$ orbifold SUSY GUTs
can result in a set of $M_{l}$ and $\overline{M}_l$ supermultiplets from extra  $27^{\prime}_l$ and $\overline{27}^{\prime}_l$ representations
\cite{Nevzorov:2012hs}. Since $M_{l}$ and $\overline{M}_l$ have opposite quantum numbers all gauge anomalies still cancel. In the case of SE$_6$SSM
the set of $M_l$ and $\overline{M}_l$ includes a pair of superfields $S$ and $\overline{S}$ as well as three pairs of $SU(2)_W$ doublets, i.e.
$H_d$ and $\overline{H}_d$, $H_u$ and $\overline{H}_u$, $L_4$ and $\overline{L}_4$. Only supermultiplets $L_4$, $\overline{L}_4$, $S$, $\overline{S}$,
$H_d$ and $H_u$ are required to be even under the $\tilde{Z}^{H}_2$ symmetry that forbids the tree-level flavor-changing transitions,
as well as the most dangerous baryon and lepton number violating operators. All other supermultiplets are expected to be odd under
this discrete symmetry\cite{Athron:2014pua}.

The $\tilde{Z}^{H}_2$ symmetry forbids all terms in the SE$_6$SSM superpotential that come from $27_i \times 27_j \times 27_k$
where $i,j,k=1,2,3$ are family indexes. Nevertheless it allows the interactions which originate from $27'_l \times 27'_m \times 27'_n$
and $27'_l \times 27_i \times 27_k$. Here indexes $l,m,n$ are associated with the supermultiplets $M_l$. As a consequence $\tilde{Z}^{H}_2$ symmetry
forbids all Yukawa interactions in $W_1$ eliminating the most dangerous operators leading to rapid proton decay. On the other hand this
symmetry allows the terms $(Q_i L_4) \overline{D}_k$ in the superpotential that permits the lightest exotic colored state (quark or squark) to decay.
In the SE$_6$SSM all charged leptons and the down-type quarks couple to just $H_d$ while the up-type quarks interact with $H_u$ only.
Thus at tree-level non-diagonal flavor transitions are suppressed.

Using the results of the analysis presented in~\cite{Hesselbach:2007te}--\cite{Hesselbach:2008vt},
it was shown that within the E$_6$SSM and its simplest modifications the lightest SUSY particles (LSPs)
are linear superpositions of the fermion components of the superfields $S_{i}$~\cite{Hall:2010ix,Hall:2010ny}.
In the simplest scenarios these states are either massless or have masses which are much smaller than $1\,\mbox{eV}$
forming hot dark matter in our Universe. The presence of very light neutral fermions may have interesting implications
for the neutrino physics (see, for example \cite{Frere:1996gb}.

To avoid the appearance of the exotic fermions with tiny masses it is assumed that the low energy matter content of the SE$_6$SSM
involves at least four $E_6$ singlet superfields. One of these superfields $\phi$ is even under the $\tilde{Z}^{H}_2$ symmetry
whereas three others $\phi_i$ are odd. The SE$_6$SSM implies that $\overline{H}_u$ and $\overline{H}_d$ get combined with the
superposition of the appropriate components from the $27_i$, composing vectorlike states with masses of order $M_X$.
At the same time the components of the supermultiplets $S$ and $\overline{S}$ as well as $L_4$ and $\overline{L }_4$ gain
the TeV scale masses. The presence of $L_4$ and $\overline{L}_4$ at low energies facilitates the gauge coupling unification \cite{King:2007uj}
and permits the lightest exotic colored state (quark or squark) to decay within a reasonable time. As a result the components
of the supermultiplets
\begin{equation}
\begin{array}{c}
(Q_i,\,u^c_i,\,d^c_i,\,L_i,\,e^c_i,\,N_i^c)
+(D_i,\,\bar{D}_i) + S_{i} + \phi_i + (H^u_{\alpha},\,H^d_{\alpha})\\[0mm]
+L_4+\overline{L}_4+S+\overline{S}+H_u+H_d+\phi\,,
\end{array}
\label{5}
\end{equation}
survive to low energies, i.e. they have masses which are many orders of magnitude smaller than $M_X$.
Here $i=1,2,3$ and $\alpha=1,2$. The $U(1)_N$ and $U(1)_{Y}$ charges of the supermultiplets listed in Equation~(\ref{5})
are given in Table~\ref{tab10}. It is worth noting that the superfields $\phi_i$, $N^c_i$ and $\phi$ do not
participate in the gauge interactions. Therefore these superfields are not included in Table~\ref{tab10}.

\begin{table}[ht]
\centering
\begin{tabular}{|c|c|c|c|c|c|c|c|c|c|c|c|c|}
\hline
& $Q_i$ & $u^c_i$ & $d^c_i$ & $L_i, L_4$ & $e^c_i$ & $S_i, S$ & $H^u_{\alpha}, H_u$ & $H^d_{\alpha}, H_d$ & $D_i$ & $\overline{D}_i$ &
$\overline{L}_4$ & $\overline{S}$\\
\hline
$\sqrt{\frac{5}{3}}Q^{Y}_i$ & $\frac{1}{6}$ & $-\frac{2}{3}$ & $\frac{1}{3}$ & $-\frac{1}{2}$ & $1$ & $0$ & $\frac{1}{2}$ & $-\frac{1}{2}$ & $-\frac{1}{3}$ & $\frac{1}{3}$ & $\frac{1}{2}$ & $0$\\
\hline
$\sqrt{{40}}Q^{N}_i$ & $1$ & $1$ & $2$ & $2$ & $1$ & $5$ & $-2$ & $-3$ & $-2$ & $-3$ & $-2$ & $-5$ \\
\hline
\end{tabular}
\caption{The $U(1)_N$ and $U(1)_{Y}$ charges of matter supermultiplets in the SE$_6$SSM.}
\label{tab10}
\end{table}

The most general renormalisable superpotential of the SE$_6$SSM, which is allowed by the $\tilde{Z}^{H}_2$,
$Z_{2}^{M}$ and $SU(3)_C \times SU(2)_W \times U(1)_{Y}\times U(1)_{N}$ symmetries, is given by
\begin{equation}
\begin{array}{c}
W_{\rm SE_6SSM} = \lambda S (H_u H_d) - \sigma \phi S \overline{S} +
\dfrac{\kappa}{3}\phi^3+\dfrac{\mu}{2}\phi^2+\Lambda\phi
+ \mu_L L_4\overline{L}_4+ \tilde{\sigma} \phi L_4\overline{L}_4 + W_{IH}\\[2mm]
+ \kappa_{ij} S (D_{i} \overline{D}_{j}) + g^D_{ij} (Q_i L_4) \overline{D}_j+ h^E_{i\alpha} e^c_{i} (H^d_{\alpha} L_4)
+ g_{ij} \phi_i \overline{L}_4 L_j + W_N + W_{\rm MSSM}(\mu=0)\,.
\end{array}
\label{6}
\end{equation}
where
\begin{equation}
W_{IH} = \tilde{M}_{ij} \phi_i \phi_j + \tilde{\kappa}_{ij} \phi \phi_i \phi_j
+ \tilde{\lambda}_{ij} \overline{S} \phi_i S_j  + \lambda_{\alpha\beta} S (H^d_{\alpha} H^u_{\beta})
+ \tilde{f}_{i\alpha} S_{i} (H^d_{\alpha} H_u) + f_{i\alpha} S_{i} (H_d H^u_{\alpha})\,,
\label{7}
\end{equation}
\begin{equation}
W_N =  \frac{1}{2} M_{ij} N_i^c N_j^c + \tilde{h}_{ij} N_i^c (H_u L_j) +
h_{i\alpha}  N_i^c (H^u_{\alpha} L_4)\,.
\label{8}
\end{equation}
In Equations~(\ref{6})--(\ref{8}) $\alpha, \beta =1,2$ and $i, j =1,2,3$ as before.
In the superpotential of the SE$_6$SSM the $U(1)_{N}$ symmetry forbids the term $\mu H_d H_u$.
However all other terms, which are present in the MSSM superpotential, are allowed.
In Equation~(\ref{6}) the sum of these terms is denoted as $W_{\rm MSSM}(\mu=0)$.
The sector responsible for the breakdown of the $SU(2)_{W}\times U(1)_{Y}\times U(1)_{N}$
symmetry involves the scalar components of $\phi$, $S$, $\overline{S}$, $H_u$ and $H_d$.
If the superfields $S$ and $\overline{S}$ develop VEVs along the D-flat direction, i.e.
$\langle S \rangle \simeq \langle \overline{S} \rangle \simeq S_0$, then the value of $S_0$
can be much larger than the sparticle mass scale $M_S$ resulting in an extremely heavy
$Z’$ boson. All extra exotic states may be also very heavy in this case.
The neutral components of $H_u$ and $H_d$ have to gain non--zero VEVs, i.e.
$\langle H_d \rangle = v_1/\sqrt{2}$ and $\langle H_u \rangle = v_2/\sqrt{2}$,
so that $v=\sqrt{v_1^2+v_2^2}\simeq 246\,\mbox{GeV}$. These VEVs generate the masses of
all SM particles triggering the breakdown of the $SU(2)_{W}\times U(1)_{Y}$ symmetry
down to $U(1)_{em}$ associated with electromagnetism. Since we further focus on the
scenarios with most sparticles at the multi-TeV scale a substantial degree of tuning is
needed to stabilize the EW scale.

For the analysis of the phenomenological implications of the SE$_6$SSM it is worth to
introduce the $Z_{2}^{E}$ symmetry, which is defined such that
$\tilde{Z}^{H}_2 = Z_{2}^{M}\times Z_{2}^{E}$ \cite{Nevzorov:2012hs}.
The supermultiplets $\overline{D}_i$, $D_i$, $H^{d}_{\alpha}$, $H^{u}_{\alpha}$, $S_{i}$, $\phi_i$
$L_4$ and $\overline{L}_4$ are odd under the $Z_{2}^{E}$ symmetry. The components of all other
supermultiplets are $Z_{2}^{E}$ even. Because the Lagrangian of the SE$_6$SSM is invariant under
both $\tilde{Z}^{H}_2$ and $Z_{2}^{M}$ symmetries, the $Z_{2}^{E}$ symmetry and
$R$--parity are also conserved. This means that the exotic states, which are odd
under the $Z_{2}^{E}$ symmetry, can only be created in pairs in collider experiments and
the lightest exotic particle as well as the lightest $R$--parity odd state have to be stable and
may contribute to the density of dark matter. Here we focus on the scenarios in which gravitino
is the lightest R-parity odd state. Recently the cosmological implications of the gravitino with
mass $m_{3/2}\sim \mbox{KeV}$ were discussed~\cite{Gu:2020ozv}. It is also assumed that
the lightest stable exotic state is predominantly formed by the fermion components of
$H^{d}_{\alpha}$ and $H^{u}_{\alpha}$.

In order to find a viable scenarios with stable gravitino one needs to ensure that the lightest unstable
R-parity odd (or exotic) state $Y$ decays before BBN, i.e. its lifetime $\tau_Y\lesssim 1\,\mbox{sec}$.
Otherwise the decay products of $Y$ can alter the abundances of light elements which are induced by the BBN.
The lifetime of the particle $Y$ decaying into gravitino and its SM partner (or the lightest
$Z_{2}^{E}$ odd state) can be estimated as \cite{Feng:2004mt}
\begin{equation}
\tau_Y \sim 48\pi \dfrac{m_{3/2}^2 M_{Pl}^2}{m_Y^5}\,,
\label{9}
\end{equation}
where $m_Y$ is its mass. For $m_Y\simeq 1\,\mbox{TeV}$ one can get $\tau_Y\lesssim 1\,\mbox{sec}$ if $m_{3/2}\lesssim 1\,\mbox{GeV}$.
When gravitinos originate from scattering of particles in the thermal bath their contribution to the dark matter density is
proportional to the reheating temperature $T_R$ \cite{Bolz:2000fu,Eberl:2020fml}
\begin{equation}
\Omega_{3/2} h^2 \sim 0.27 \left(\dfrac{T_R}{10^8 GeV}\right) \left(\dfrac{1\,\mbox{GeV}}{m _{3/2}}\right) \left(\dfrac{M_{\tilde{g}}}{1\,\mbox{TeV}}\right)^2\,.
\label{10}
\end{equation}
In Equation~(\ref{10}) $M_{\tilde{g}}$ is a gluino mass. Since $\Omega_{3/2} h^2 \le 0.12$ \cite{Ade:2015xua},
for $m _{3/2}\simeq 1\,\mbox{GeV}$ and $M_{\tilde{g}}\gtrsim 3\,\mbox{TeV}$ one finds an upper bound
$T_R\lesssim 10^{6-7}\,\mbox{GeV}$ \cite{Hook:2018sai}.

\section{Generation of lepton and baryon asymmetries}

Even for so low reheating temperatures the appropriate amount of the lepton asymmetry can be induced
within the SE$_6$SSM via the out--of equilibrium decays of the lightest right-handed neutrino/sneutrino.
Due to $(B+L)$--violating sphaleron interactions the generated lepton asymmetry is converted into the baryon asymmetry.

In the SM the process of the generation of lepton asymmetry is controlled by the
three flavor CP (decay) asymmetries $\varepsilon_{1,\,\ell_k}$ which are associated with three lepton flavors.
These decay asymmetries appear on the right--hand side of Boltzmann equations. They are defined as
\begin{equation}
\varepsilon_{1,\,\ell_k}=\dfrac{\Gamma_{N_1 \ell_{k}}-\Gamma_{N_1 \bar{\ell}_{k}}}
{\sum_{m} \left(\Gamma_{N_1 \ell_{m}}+\Gamma_{N_1 \bar{\ell}_{m}}\right)}\,.
\label{11}
\end{equation}
Here $\Gamma_{N_1 \ell_{k}}$ and $\Gamma_{N_1 \bar{\ell}_{k}}$ are partial widths of the lightest right-handed neutrino decays
$N_1\to L_k+H_u$ and $N_1\to \overline{L}_k+H^{*}_u$ with $k,m=1,2,3$. At the tree level $\Gamma_{N_1 \ell_{k}}=\Gamma_{N_1 \bar{\ell}_{k}}$
and CP asymmetries (\ref{11}) vanish. The non--zero contributions to the decay asymmetries come from the interference between the tree--level
amplitudes of the decays of $N_1$ and one--loop corrections to them if CP invariance is violated in the lepton sector.

In the MSSM the decays of the lightest right--handed neutrino into Higgsino $\widetilde{H}_u$ and sleptons $\widetilde{L}_k$
also contribute to the lepton asymmetry generation. The corresponding flavor decay asymmetries are given by
\begin{equation}
\varepsilon_{1,\,\widetilde{\ell}_k}=\dfrac{\Gamma_{N_1 \widetilde{\ell}_{k}}-\Gamma_{N_1 \widetilde{\ell}^{*}_{k}}}
{\sum_{m} \left(\Gamma_{N_1 \widetilde{\ell}_{m}}+\Gamma_{N_1 \widetilde{\ell}^{*}_{m}}\right)}\,.
\label{12}
\end{equation}
Moreover supersymmetry predicts the existence of the lightest right--handed sneutrino $\widetilde{N}_1$ which is
a scalar partner of $N_1$. The decays of $\widetilde{N}_1$ into slepton and Higgs as well as into lepton and Higgsino
provide another possible origin of lepton asymmetry. The corresponding decay asymmetries can be determined similarly
to the neutrino ones
\begin{equation}
\varepsilon_{\widetilde{1},\,\ell_k}=\dfrac{\Gamma_{\widetilde{N}_1^{*} \ell_{k}}-\Gamma_{\widetilde{N}_1 \bar{\ell}_{k}}}
{\sum_{m} \left(\Gamma_{\widetilde{N}_1^{*} \ell_{m}}+\Gamma_{\widetilde{N}_1 \bar{\ell}_{m}}\right)}\,,\qquad
\varepsilon_{\widetilde{1},\,\widetilde{\ell}_k}=\dfrac{\Gamma_{\widetilde{N}_1 \widetilde{\ell}_{k}}-\Gamma_{\widetilde{N}_1^{*}
\widetilde{\ell}^{*}_{k}}}
{\sum_{m} \left(\Gamma_{\widetilde{N}_1 \widetilde{\ell}_{m}}+\Gamma_{\widetilde{N}_1^{*} \widetilde{\ell}^{*}_{m}}\right)}\,.
\label{13}
\end{equation}
When the sparticle mass scale $M_S$ is considerably smaller than $M_1$
\begin{equation}
\varepsilon_{1,\,\ell_k}=\varepsilon_{1,\,\widetilde{\ell}_k}=\varepsilon_{\widetilde{1},\,\ell_k}=
\varepsilon_{\widetilde{1},\,\widetilde{\ell}_k}\,.
\label{14}
\end{equation}

Assuming the type I seesaw models of neutrino mass generation
the decay asymmetries mentioned above were initially computed within the SM \cite{CPasym-SM-1}--\cite{CPasym-SM-4} and
MSSM \cite{CPasym-SUSY-1}--\cite{CPasym-SUSY-3}. Flavor effects were ignored in the early studies of leptogenesis (see for
example \cite{Buchmuller:2004nz}). The importance of these effects was emphasised in \cite{lg-flav-1}--\cite{Antusch:2006cw}.

The non-minimal SUSY models (like the SE$_6$SSM) may include additional $SU(2)_W$ doublets with quantum numbers of Higgs
fields ($H^{d}_{\alpha}$ and $H^{u}_{\alpha}$) and extra lepton multiplets ($L_4$ and $\overline{L}_4$) at low energies.
It is convenient to denote all Higgs like multiplets and $SU(2)_W$ lepton doublets, that interact with the right--handed neutrino
superfields, as $H^u_k$ and $L_x$ respectively. In the case of the SE$_6$SSM $H^u_3\equiv H_u$, $k=1,2,3$ and $x=1,2,3,4$.
If the components of additional Higgs like and lepton supermultiplets are lighter than $N_1$ and $\widetilde{N}_1$ they can give rise
to new decay modes of the lightest right--handed neutrino and its superpartner. Each new channel of the decays of $N_1$ and
$\widetilde{N}_1$ should lead to extra CP asymmetry that contributes to the lepton asymmetry generation. In this case the
definitions of the decay asymmetries (\ref{11})--(\ref{13}) need to be generalised. In particular the definitions (\ref{11})--(\ref{12})
can be modified in the following way \cite{Nevzorov:2017gir}
\begin{equation}
\varepsilon^{k}_{1,\,f}=\dfrac{\Gamma^{k}_{N_1 f}-\Gamma^{k}_{N_1 \bar{f}}}
{\sum_{m,\,f'} \left(\Gamma^{m}_{N_1 f'}+\Gamma^{m}_{N_1 \bar{f}'}\right)}\,,
\label{15}
\end{equation}
where $f$ and $f'$ may be either $\ell_x$ or $\widetilde{\ell}_x$ while $\bar{f}$ and $\bar{f}'$ should be associated
with either $\bar{\ell}_x$ or $\widetilde{\ell}_x^{*}$. The superscripts $k$ and $m$ correspond to the components of the
supermultiplets $H^{u}_{k}$ and $H^{u}_{m}$ in the final state. The denominator of Equation~(\ref{15}) contains a sum of
partial widths of the decays of $N_1$. For $\varepsilon^{k}_{1,\,\ell_x}$ this sum involves all possible partial decay
widths of the lightest right--handed neutrino whose final state includes fermion components of the supermultiplets $L_x$.
The expressions for $\varepsilon^{k}_{1,\,\widetilde{\ell}_x}$ involve in the denominator a sum of partial widths
of the decays of the lightest right--handed neutrino over all possible decay modes which have scalar components of
the supermultiplets $L_x$ in the final state.

The CP asymmetries associated with the decays of the lightest right--handed sneutrino $\varepsilon^{k}_{\widetilde{1},\,f}$
can be defined similarly to $\varepsilon^{k}_{1,\,f}$. In order to get the appropriate expressions for $\varepsilon^{k}_{\widetilde{1},\,f}$
the field of the lightest right--handed neutrino in Equation~(\ref{15}) ought to be replaced by either $\widetilde{N}_1$ or $\widetilde{N}_1^{*}$.
In the limit, when the sparticle mass scale $M_S$ is negligibly small as compared with $M_1$, all soft SUSY breaking terms can be safely ignored
and the relation between different decay asymmetries (\ref{14}) remains intact, i.e. $\varepsilon^{k}_{1,\,f}=\varepsilon^{k}_{\widetilde{1},\,f}$.

Within the SE$_6$SSM $\varepsilon^{3}_{1,\,\ell_n}$ ($\varepsilon^{3}_{1,\,\widetilde{\ell}_n}$) with $n=1,2,3$ are flavor
decay asymmetries associated with the decays of $N_1$ into Higgs doublet $H_u$ and ordinary leptons (Higgsino $\widetilde{H}_u$ and sleptons),
whereas $\varepsilon^{3}_{\widetilde{1},\,\ell_n}$ ($\varepsilon^{3}_{\widetilde{1},\,\widetilde{\ell}_n}$) are CP asymmetries
corresponding to the decays of $\widetilde{N}_1$ into leptons and Higgsino $\widetilde{H}_u$ (sleptons and Higgs doublet $H_u$).
Additional decay asymmetries $\varepsilon^{\alpha}_{1,\,\ell_4}$, $\varepsilon^{\alpha}_{1,\,\widetilde{\ell}_4}$, $\varepsilon^{\alpha}_{\widetilde{1},\,\ell_4}$ and $\varepsilon^{\alpha}_{\widetilde{1},\,\widetilde{\ell}_4}$ in this SUSY model
arise due to the extra decay channels of $N_1$ and $\widetilde{N}_1$
\begin{equation}
N_1\to L_4 + H^u_{\alpha},\qquad N_1\to \widetilde{L}_4+\widetilde{H}^u_{\alpha},\qquad
\widetilde{N}^{*}_1\to L_4 + \widetilde{H}^{u}_{\alpha},\qquad \widetilde{N}_1\to \widetilde{L}_4+ H^u_{\alpha},
\label{16}
\end{equation}
The structure of the part of the SE$_6$SSM superpotential $W_N$ (\ref{8}), that describes the interactions of the right--handed neutrino
superfields with other supermultiplets, indicates that all other $\varepsilon^{k}_{1,\,f}$ and $\varepsilon^{k}_{\widetilde{1},\,f}$ vanish.

After inflation the lightest right-handed neutrino/sneutrino with mass $M_1$ may be produced by thermal scattering if $T_R > M_1$.
Since in the scenarios under consideration $T_R\lesssim 10^{6-7}\,\mbox{GeV}$ here we require that $M_1\lesssim 10^6\,\mbox{GeV}$
to guarantee that thermal leptogenesis can take place. It is also assumed that two other right--handed neutrino/sneutrino states
have masses $M_{2,3}\lesssim 10^{6}\,\mbox{GeV}$ so that $M_1 \lesssim M_2 \lesssim M_3$ while the sparticle mass scale $M_S$
is lower than $10\,\mbox{TeV}$. In order to reproduce the left--handed neutrino mass scale $m_{\nu}\lesssim 0.1\,\mbox{eV}$ the
absolute values of the couplings $|\tilde{h}_{ij}|$ in Equation~(\ref{8}) should be rather small for so low $M_i$,
i.e. $|\tilde{h}_{ij}|^2\ll 10^{-8}$. Such couplings $\tilde{h}_{ij}$ induce quite small decay asymmetries
$\varepsilon^{3}_{1,\,f}$ and $\varepsilon^{3}_{\widetilde{1},\,f}$ so that $\tilde{h}_{ij}$ can be ignored.

Nevertheless the new decay modes of $N_1$ and $\widetilde{N}_1$ (\ref{16}) may give rise to the sufficiently large CP asymmetries
$\varepsilon^{\alpha}_{1,\,\ell_4}$, $\varepsilon^{\alpha}_{1,\,\widetilde{\ell}_4}$, $\varepsilon^{\alpha}_{\widetilde{1},\,\ell_4}$ and $\varepsilon^{\alpha}_{\widetilde{1},\,\widetilde{\ell}_4}$ that control the process of lepton asymmetry generation.
At the tree level the partial widths corresponding to the new decay channels (\ref{16}) are given by
\begin{equation}
\Gamma^{\alpha}_{N_1 \ell_4}+\Gamma^{\alpha}_{N_1 \bar{\ell}_4}=\Gamma^{\alpha}_{N_1 \widetilde{\ell}_4}+\Gamma^{\alpha}_{N_1 \widetilde{\ell}^{*}_4}=
\Gamma^{\alpha}_{\widetilde{N}_1^{*}\ell_4}=\Gamma^{\alpha}_{\widetilde{N}_1 \bar{\ell}_4}=\Gamma^{\alpha}_{\widetilde{N}_1 \widetilde{\ell}_4}=
\Gamma^{\alpha}_{\widetilde{N}_1^{*} \widetilde{\ell}_4^{*}}=\frac{|h_{1\alpha}|^2}{8\pi}M_1
\label{17}
\end{equation}
and CP asymmetries $\varepsilon^{\alpha}_{1,\,\ell_4}$, $\varepsilon^{\alpha}_{1,\,\widetilde{\ell}_4}$,
$\varepsilon^{\alpha}_{\widetilde{1},\,\ell_4}$ as well as $\varepsilon^{\alpha}_{\widetilde{1},\,\widetilde{\ell}_4}$ vanish.
As in the SM and MSSM the non--zero values of the decay asymmetries in the SE$_6$SSM arise after the
inclusion of one--loop self--energy and vertex corrections to the decay amplitudes of the lightest right--handed neutrino/sneutrino.
Neglecting the Yukawa couplings $\tilde{h}_{ik}$ one finds \cite{Nevzorov:2017gir}
\begin{equation}
\begin{array}{c}
\varepsilon^{\alpha}_{1,\,\ell_4}=\varepsilon^{\alpha}_{1,\,\widetilde{\ell}_4}=\varepsilon^{\alpha}_{\widetilde{1},\,\ell_4}=
\varepsilon^{\alpha}_{\widetilde{1},\,\widetilde{\ell}_4}=\dfrac{1}{8\pi}\dfrac{\sum_{j=2,3}
\mbox{Im}\biggl[h^{*}_{1\alpha} B_{j} h_{j\alpha} \biggr]}{\sum_{\beta} |h_{1\beta}|^2}\,,\\[1mm]
B_{j}=\sum_{\beta}\left\{h^{*}_{1\beta} h_{j\beta}\, f\left(\dfrac{M^2_j}{M_1^2}\right)+
\dfrac{M_1}{M_j} h_{1\beta} h^{*}_{j\beta}f^S\left(\dfrac{M^2_j}{M_1^2}\right)\right \}\,,\\[1mm]
f(z)=f^{V}(z)+f^{S}(z)\,,\qquad f^S(z)=\dfrac{2\sqrt{z}}{1-z}\,,\qquad f^V(z)=-\sqrt{z}\,\ln\left(\dfrac{1+z}{z}\right)\,.
\end{array}
\label{18}
\end{equation}

The analytical expressions for the CP asymmetries (\ref{18}) are simplified dramatically if $|h_{12}|$ goes to zero.
In this case $\varepsilon^{2}_{1,\,\ell_4}=\varepsilon^{2}_{1,\,\widetilde{\ell}_4}=
\varepsilon^{2}_{\widetilde{1},\,\ell_4}=\varepsilon^{2}_{\widetilde{1},\,\widetilde{\ell}_4}=0$.
If $M_j$ are real and $h_{j1}=|h_{j1}| e^{i\varphi_{j1}}$ the expressions for other CP asymmetries reduce to \cite{Nevzorov:2017gir,Nevzorov:2018leq}
\begin{equation}
\varepsilon^{1}_{1,\,\ell_4}=\varepsilon^{1}_{1,\,\widetilde{\ell}_4}=
\varepsilon^{1}_{\widetilde{1},\,\ell_4}=\varepsilon^{1}_{\widetilde{1},\,\widetilde{\ell}_4}=\varepsilon=
\frac{1}{8\pi}\Biggl[\sum_{j=2,3} |h_{j1}|^2 f\left(\frac{M^2_j}{M_1^2}\right) \sin 2\Delta\varphi_{j1}\Biggr]\,,
\label{19}
\end{equation}
where $\Delta\varphi_{j1}=\varphi_{j1}-\varphi_{11}$\,. From the part of the SE$_6$SSM superpotential (\ref{8})
one can see the supermultiplets $H^{u}_{\alpha}$ may be redefined so that only $H^{u}_{1}$ interacts with $N^c_1$ and $L_4$.
Thus $\tilde{h}_{12}$ in $W_N$ may be set to zero without loss of generality. It is also worth noting that
the scalar and fermion components of the supermultiplet $L_4$ being produced in the decays of $N_1$ and $\widetilde{N}_1$
sequentially decay into ordinary leptons inducing lepton number asymmetries.

The evolution of the $U(1)_{B-L}$ number densities is described by the Boltzmann equations.
In the scenarios under consideration the results for the baryon and lepton asymmetries obtained within the SM and MSSM
can be easily generalised. In particular, the generated total baryon asymmetry may be estimated using an approximate
formula \cite{Davidson:2008bu}:
\begin{equation}
Y_{\Delta B}\sim 10^{-3} \varepsilon \eta \,,\qquad\qquad
Y_{\Delta B}=\dfrac{n_B-n_{\bar{B}}}{s}\biggl|_0=(8.75\pm 0.23)\times 10^{-11}\,,
\label{20}
\end{equation}
where $Y_{\Delta B}$ is the baryon asymmetry relative to the entropy density.
In Equation~(\ref{20}) $\eta$ is an efficiency factor that varies from 0 to 1.
The efficiency factor in the so--called strong washout scenario can be estimated as follows
\begin{equation}
\begin{array}{c}
\eta \simeq H(T=M_1)/\Gamma_{1}\,,\\[1mm]
\Gamma_1 = \Gamma^{1}_{N_1 \ell_4}+\Gamma^{1}_{N_1 \bar{\ell}_4}=\dfrac{|h_{11}|^2}{8\pi}\,M_1\,,
\qquad\qquad
H=1.66 g_{*}^{1/2}\dfrac{T^2}{M_{P}}\,.
\end{array}
\label{21}
\end{equation}
Here $M_{P}=1.22\cdot 10^{19}\,\mbox{GeV}$, $H$ is the Hubble expansion rate and $g_{*}=n_b+\dfrac{7}{8}\,n_f$ is the number of relativistic
degrees of freedom in the thermal bath. In the SE$_6$SSM $g_{*}\approx 360$.

\begin{figure}[h!]
\centering
\includegraphics[width=6.8cm]{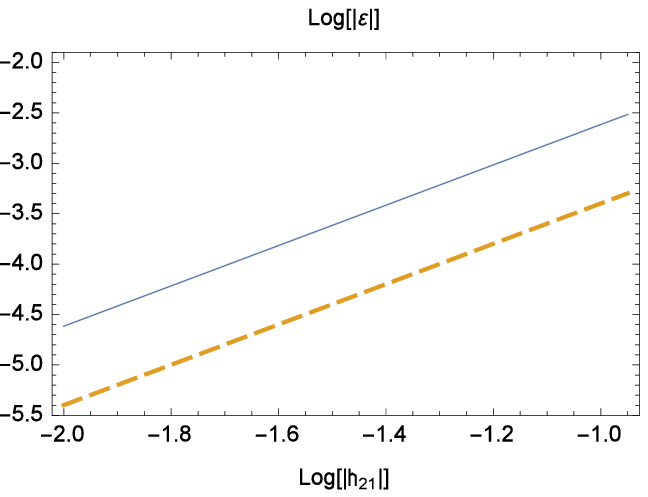}\qquad
\includegraphics[width=6.8cm]{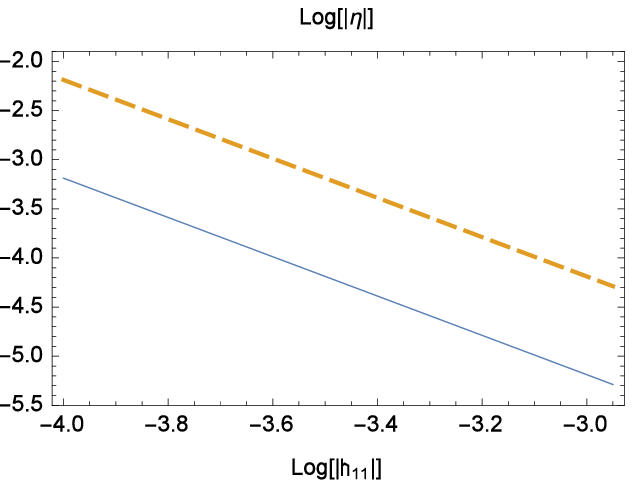}
\caption{({\bf Left}) Logarithm (base 10) of the absolute value of the CP asymmetry $|\varepsilon|$ as a function of logarithm (base 10)
of $|h_{21}|$ for $M_2=1.2\cdot M_1$ (solid line) and $M_2=10\cdot M_1$ (dashed line). ({\bf Right}) Logarithm (base 10) of the absolute
value of the efficiency factor $|\eta|$ as a function of logarithm (base 10) of $|h_{11}|$ for $M_1=100\,\mbox{TeV}$ (solid line) and
$M_1=1000\,\mbox{TeV}$ (dashed line). Here we fix $|h_{12}|=|h_{31}|=0$ and $\Delta\varphi_{21}=\pi/4$.}
\label{fig5}
\end{figure}
\unskip

In order to simplify our numerical analysis we set $|h_{12}|=|h_{31}|=0$. Our results are summarised in Figure~\ref{fig5}.
The decay asymmetries (\ref{19}) are determined by $|h_{21}|$, $(M_2/M_1)$ and combination of the CP--violating phases
$\Delta\varphi_{21}$, but do not depend on $|h_{11}|$. Here we fix $\Delta\varphi_{21}$ so that these asymmetries attain their
maximum absolute values, i.e. $|\sin 2\Delta\varphi_{21}|=1$. In this case $|\varepsilon|$ changes from $2.4\cdot 10^{-5}$
to $2.4\cdot 10^{-3}$ if $|h_{21}|$ increases from $0.01$ to $0.1$ for $M_2=1.2\cdot M_1$\,.
As follows from Equation~(\ref{21}) the efficiency factor $\eta$ is set by $|h_{11}|$ and $M_1$. We restrict our consideration
here by the values of $|h_{11}|^2 \gg |\tilde{h}_{ik}|^2$, i.e. $|h_{11}| \gtrsim 10^{-4}$. For $M_1=100\,\mbox{TeV}$ the
efficiency factor varies from $6.5\cdot 10^{-4}$ to $6.5\cdot 10^{-6}$ when $|h_{11}|$ increases from $10^{-4}$ to $10^{-3}$.
Thus for $|h_{21}|\simeq 0.1$, $M_1=100\,\mbox{TeV}$ and $M_2=1.2\cdot M_1$ the observed baryon asymmetry can be reproduced
even if $|h_{11}|\gtrsim 10^{-4}$\,.

\section{Dark matter-nucleon scattering cross section}

The scalar components of the supermultiplets $\phi_i$, $S_i$, $H^u_{\alpha}$ and $H^d_{\alpha}$ do not acquire VEVs.
Their fermion components form the exotic (inert) neutralino and chargino states. The signatures associated
with the inert neutralino states were examined in \cite{Khalil:2021afa,Khalil:2021tpz}.
When the components of $\phi_i$ are significantly heavier than the fermions and bosons from $S_i$, $H^u_{\alpha}$ and $H^d_{\alpha}$,
they can be integrated out so that $W_{IH}$ reduces to
\begin{equation}
\begin{array}{c}
W_{IH} \to \widetilde{W}_{IH}\simeq -\widetilde{\mu}_{ij} S_i S_j + \lambda_{\alpha\beta} S (H^d_{\alpha} H^u_{\beta})
+ \tilde{f}_{i\alpha} S_{i} (H^d_{\alpha} H_u) + f_{i\alpha} S_{i} (H_d H^u_{\alpha})+...\,.
\end{array}
\label{22}
\end{equation}
Here and further we work in a field basis in which $\widetilde{\mu}_{ij}=\widetilde{\mu}_{i}\,\delta_{ij}$ and
$\lambda_{\alpha\beta}=\lambda_{\alpha\alpha}\,\delta_{\alpha\beta}$.

In this article we explore the scenarios in which the fermion components of the $H^u_{1}$ and $H^d_{1}$
compose the lightest exotic state with $Z_{2}^{E}=-1$ while all other exotic states and all sparticles except gravitino
have masses which are considerably larger than $1\,\mbox{TeV}$. We also assume that $H^d_{1}$ and $H^u_{1}$
mostly interact with $S_1$, $H_u$ and $H_d$, whereas all other couplings of the supermultiplets $H^u_{1}$ and $H^d_{1}$
are very small. In this approximation the mass matrix, that determines the lightest exotic neutralino masses,
takes the form \cite{Nevzorov:2022zns}
\begin{equation}
M^{ab}=-
\left(
\begin{array}{ccc}
0                                           & \mu_{11}                            & \dfrac{\tilde{f}_{11}}{\sqrt{2}} v_2 \\[3mm]
\mu_{11}                                    & 0                                   & \dfrac{f_{11}}{\sqrt{2}} v_1 \\[3mm]
\dfrac{\tilde{f}_{11}}{\sqrt{2}} v_2        & \dfrac{f_{11}}{\sqrt{2}} v_1        & \widetilde{\mu}_1             \\
\end{array}
\right)\,,
\label{23}
\end{equation}
where $\mu_{11}\simeq \lambda_{11} \langle S \rangle$. Instead of the VEVs of $H_d$ and $H_u$, i.e.
$v_1$ and $v_2$, it is more convenient to introduce $\tan\beta=v_2/v_1$ and $v=\sqrt{v_1^2+v_2^2} \approx 246\,\mbox{GeV}$.
The charged fermion components of the supermultiplets $H^u_{1}$ and $H^d_{1}$ form the lightest exotic chargino.
Its mass is determined by $\mu_{11}$, i.e. $m_{\chi^{\pm}_1}=|\mu_{11}|$.

If $|\widetilde{\mu}_1|$ is considerably larger than $|\mu_{11}|$ and $v$ the mass matrix (\ref{23}) can be diagonalised.
Using the perturbation theory method (see, for example, \cite{Kovalenko:1998dc}--\cite{Nevzorov:2004ge}), one finds \cite{Nevzorov:2022zns}
\begin{equation}
\begin{array}{c}
m_{\chi_1} \simeq m_{\chi^{\pm}_1} - \Delta_1\,,\qquad\quad m_{\chi_2} \simeq m_{\chi^{\pm}_1} + \Delta_2\,,\qquad\quad
m_{\chi_3} \simeq \widetilde{\mu}_1 + \Delta_1 + \Delta_2\,,\\[2mm]
\Delta_1 \simeq \dfrac{(\tilde{f}_{11} v\sin\beta + f_{11} v\cos\beta)^2}{4(\widetilde{\mu}_1-m_{\chi^{\pm}_1})}\,,\qquad\qquad
\Delta_2 \simeq \dfrac{(\tilde{f}_{11} v\sin\beta - f_{11} v\cos\beta)^2}{4(\widetilde{\mu}_1+m_{\chi^{\pm}_1})}\,.
\end{array}
\label{24}
\end{equation}
As follows from Equation~(\ref{24}) the lightest exotic neutralino masses ($m_{\chi_2}$ and $m_{\chi_1}$) are also set by $\mu_{11}$
in the leading approximation. We restrict our consideration here to the part of the parameter space of the SE$_6$SSM
where $m_{\chi_2}-m_{\chi_1}> 200\,\mbox{MeV}$. As a consequence the inelastic scattering processes $\chi_1 N\to \chi_2 N$,
where $N$, $\chi_1$ and $\chi_2$ denote a nucleon as well as the lightest and second lightest exotic neutralino states,
do not take place. In this part of the parameter space $\chi_2$ decays before BBN, i.e. its lifetime is shorter than $1\,\mbox{s}$\,.

Since in the scenarios under consideration the lightest neutral exotic neutralino $\chi_1$ is stable its contribution to the cold dark
matter density may be estimated using formula
\begin{equation}
\Omega_{\tilde{H}} h^2 \simeq 0.1 \, \left(\dfrac{\mu_{11}}{1\,\mbox{TeV}}\right)^2\,.
\label{25}
\end{equation}
The approximate formula (\ref{25}) was derived within the MSSM \cite{Arkani-Hamed:2006wnf,Chalons:2012xf}.
On the other hand the Planck observations lead to $(\Omega h^2)_{\text{exp}} = 0.1188 \pm 0.0010$ \cite{Ade:2015xua}.
Therefore in the phenomenologically viable scenarios $\mu_{11}$ should be lower than $1.1\,\mbox{TeV}$.
When $\mu_{11} < 1.1\,\mbox{TeV}$, gravitino can account for some or major part of the cold
dark matter density.

In the SE$_6$SSM the interactions of the cold dark matter with the SM particles are determined
by the couplings of $\chi_1$ because the corresponding gravitino couplings are negligibly small.
The low energy effective Lagrangian, that describes the interactions of the lightest exotic neutralino with quarks
can be written as
\begin{equation}
\mathcal{L}_{\chi_1 q}=\sum_q \Bigl( a_q \bar{\chi}_1\chi_1 \bar{q}q + d_q \bar{\chi}_1 \gamma^{\mu}\gamma_5 \chi_1 \bar{q} \gamma_{\mu}\gamma_5 q\Bigr)\,.
\label{26}
\end{equation}
The first term in the brackets results in a spin--independent interaction whereas the second one
gives rise to a spin--dependent interaction.

In the scenarios under consideration the dominant contribution to the parameters $d_q$ in the Lagrangian~(\ref{26})
stems from $t$--channel $Z$ boson exchange. Taking into account that in the field basis $(\tilde{H}^{d0}_1,\,\tilde{H}^{u0}_1,\,\tilde{S}_1)$
\begin{equation}
\chi_{\alpha} = N_{\alpha}^1 \tilde{H}^{d0}_1 + N_{\alpha}^2 \tilde{H}^{u0}_1 + N_{\alpha}^3 \tilde{S}_1\,,\qquad\qquad\qquad \alpha=1,2\,,
\label{27}
\end{equation}
where $N^a_i$ is the exotic neutralino mixing matrix defined by
\begin{equation}
N_i^a M^{ab} N_j^b = m_i \delta_{ij}, \qquad\qquad\qquad\mbox{ no sum on } i\,,
\label{28}
\end{equation}
the part of the Lagrangian, which describes the interactions of the lightest and second lightest exotic neutralino states
with $Z$, may be presented in the following form:
\begin{equation}
\mathcal{L}_{Z\chi\chi}=\sum_{\alpha,\beta}\dfrac{M_Z}{2 v}Z_{\mu}
\biggl(\chi^{T}_{\alpha}\gamma_{\mu}\gamma_{5}\chi_{\beta}\biggr) R_{Z\alpha\beta}\,,\qquad\qquad
R_{Z\alpha\beta}=N_{\alpha}^1 N_{\beta}^1 - N_{\alpha}^2 N_{\beta}^2\,.
\label{29}
\end{equation}
In Equation~(\ref{28}) $M^{ab}$ is $3\times 3$ mass matrix (\ref{23}).
Then the parameters $d_q$ as well as the corresponding $\chi_1$--proton and $\chi_1$--neutron
scattering cross sections ($\sigma^{p}$ and $\sigma^{n}$) are given by
\begin{equation}
\sigma^{p,n}=\dfrac{12 m^2_r}{\pi}\Bigg(\sum_{q=u,d,s} d_q \Delta^{p,n}_q\Biggr)^2\,,\qquad
d_q=\dfrac{T_{3q}}{2v^2}R_{Z11}\,, \qquad m_r = \dfrac{m_{\chi_1} m_N}{m_{\chi_1} + m_N}\,.
\label{30}
\end{equation}
Here $m_N$ is a nucleon mass and $T_{3q}$ is the third component of isospin. We set $\Delta^{p}_u=\Delta^{n}_d=0.842$,
$\Delta^{p}_d=\Delta^{n}_u=-0.427$ and $\Delta^{p}_s=\Delta^{n}_s=-0.085$ \cite{Chalons:2012xf}.

In the SE$_6$SSM $\chi_1$ does not couple to squarks and quarks. As a consequence the only contributions that
parameters $a_q$ receive, come from the $t$--channel exchange of Higgs scalars. Since in the scenarios under consideration
all Higgs bosons except the lightest Higgs scalar $h_1$ are expected to be considerably heavier than $1\,\mbox{TeV}$,
all contributions caused by the heavy Higgs exchange can be neglected.
The lightest Higgs boson with mass $m_{h_1}\approx 125\,\mbox{GeV}$ manifests itself in the interactions with the
SM states as a SM--like Higgs in this case so that
\begin{equation}
\dfrac{a_q}{m_q} \simeq \dfrac{g_{h\chi\chi}}{v m^2_{h_1}}\,,\qquad\qquad
g_{h\chi\chi}= -\dfrac{1}{\sqrt{2}}\Biggl(f_{11} N^3_{1} N^2_{1} \cos\beta +
\tilde{f}_{11} N^3_{1} N^1_{1} \sin\beta \Biggr)\,,
\label{31}
\end{equation}
where $m_q$ is a quark mass and $g_{h\chi\chi}$ is the coupling of the lightest exotic neutralino to $h_1$.
The spin--independent part of $\chi_1$--nucleon cross section takes the form \cite{Ellis:2008hf,Kalinowski:2008iq}
\begin{equation}
\begin{array}{c}
\sigma_{SI} = \dfrac{4 m^2_r m_N^2}{\pi v^2 m^4_{h_1}} |g_{h \chi \chi} F^N|^2 \,,\qquad
F^N = \sum_{q=u,d,s} f^N_{Tq} + \frac{2}{27} \sum_{Q=c,b,t} f^N_{TQ}\,,\\
m_N f^N_{Tq} = \langle N | m_{q}\bar{q}q |N \rangle\,,\qquad\qquad\qquad f^N_{TQ} = 1 - \sum_{q=u,d,s} f^N_{Tq}\,.
\end{array}
\label{32}
\end{equation}

The value of $\sigma_{SI}$ depends quite strongly on $f^N_{Tq}$, i.e. hadronic matrix elements.
We fix $f^N_{Ts}\simeq 0.0447$, $f^N_{Td}\simeq 0.0191$ and $f^N_{Tu}\simeq 0.0153$ which are
the default values used in micrOMEGAs \cite{Belanger:2013oya} (see also \cite{Alarcon:2011zs}--\cite{Alarcon:2012nr}).
Using the perturbation theory method it is straightforward to obtain the approximate expressions for
$g_{h\chi\chi}$ and $R_{Z11}$. If $\widetilde{\mu}_1\gg \mu_{11}>0$ and $\mu_{11}$ is substantially larger than
$f_{11} v \cos\beta$ and $\tilde{f}_{11} v \sin\beta$, one finds
\begin{equation}
|g_{h\chi\chi}|\simeq \dfrac{\Delta_1}{v}\,,\qquad\qquad
R_{Z11}\simeq \dfrac{v^2(f_{11}^2\cos^2\beta - \tilde{f}_{11}^2 \sin^2\beta)}{4 \mu_{11} (\widetilde{\mu}_1 - \mu_{11})}\,.
\label{33}
\end{equation}

In our analysis we restrict our considerations to moderate values of $\tan\beta$, i.e. $\tan\beta\simeq 2$.
For $\tan\beta\le 4$ one can get $m_{h_1}\approx 125\,\mbox{GeV}$ in the SE$_6$SSM only if $\lambda\gtrsim \sqrt{2} (M_Z/v)\simeq 0.5$.
When coupling $\lambda$ is so large all Higgs states except the SM--like Higgs boson have masses beyond the multi-TeV
range \cite{King:2005jy}--\cite{King:2005my}, \cite{King:2006vu}--\cite{King:2006rh}. Therefore they cannot be observed at the LHC.
The realisation of such scenarios requires a significant fine--tuning, $\sim 0.01\%$, of the parameters of the model under consideration \cite{Maniatis:2009re}.

LHC experiments ruled out the $U(1)_{N}$ gauge boson with masses $M_{Z'}$ below $4.5\,\mbox{TeV}$ \cite{CMS:2021ctt,ATLAS:2019erb}.
If $\langle S \rangle \simeq \langle \overline{S} \rangle$ the mass of the $Z'$ boson in the SE$_6$SSM is given by
\begin{equation}
M_{Z'}\approx 2 g^{\prime}_1 Q_S \,\langle S \rangle\,,
\label{34}
\end{equation}
where $g^{\prime}_1$ and $Q_{S}$ are the $U(1)_N$ gauge coupling and the $U(1)_N$ charge of the superfield $S$.
The low--energy value of $g^{\prime}_1$ can be calculated assuming the unification of gauge couplings \cite{King:2005jy}.
Then $M_{Z'}\gtrsim 4.5\,\mbox{TeV}$ can be obtained when $\langle S \rangle \simeq \langle \overline{S} \rangle \gtrsim 6\,\mbox{TeV}$.
To avoid the lower experimental bound on the lightest exotic chargino mass we assume that $\mu_{11}\gtrsim 200\,\mbox{GeV}$.
To ensure that $\chi_1$ leads to the phenomenologically acceptable density of the cold dark matter the interval of variations of $\mu_{11}$
is limited from above by $1\,\mbox{TeV}$ so that $\lambda_{11}\lesssim 0.17$. In addition, the validity of perturbation theory up to the
GUT scale is required that constrains the range of variations of $f_{11}$ and $\tilde{f}_{11}$ at low energies.
We also set $\widetilde{\mu}_1\simeq 2\,\mbox{TeV}$.

\begin{figure}[h!]
\centering
\includegraphics[width=6.9cm]{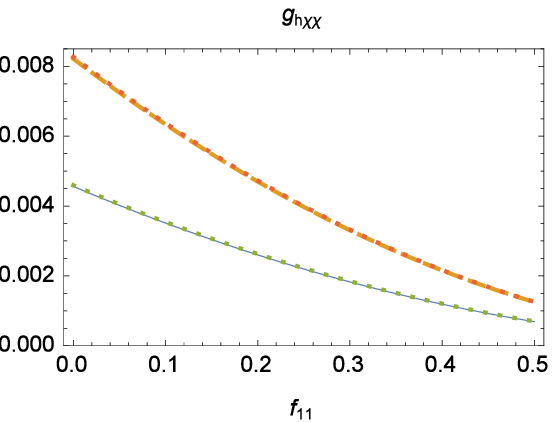}\qquad
\includegraphics[width=6.7cm]{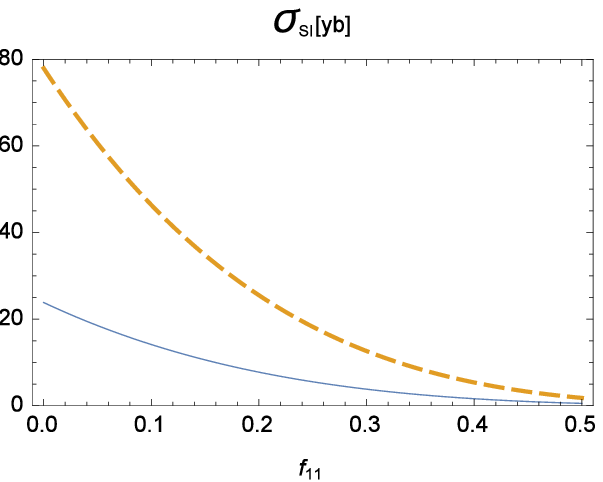}
\caption{({\bf Left})~ The~ coupling~ $g_{h\chi\chi}$~ and~ ({\bf Right}) the cross--section $\sigma_{SI}$
as a function of $f_{11}$ for $\tilde{f}_{11}=-0.41$, $\tan\beta=2$, $\widetilde{\mu}_1=2\,\mbox{TeV}$,
$\mu_{11} = 200\,\mbox{GeV}$ (solid lines) and $\mu_{11} = 1\,\mbox{TeV}$ (dashed lines). The dotted
lines correspond to the approximate expression for $g_{h\chi\chi}$ (\ref{33}).}
\label{fig10}
\end{figure}
\unskip

\begin{figure}[h!]
\centering
\includegraphics[width=6.8cm]{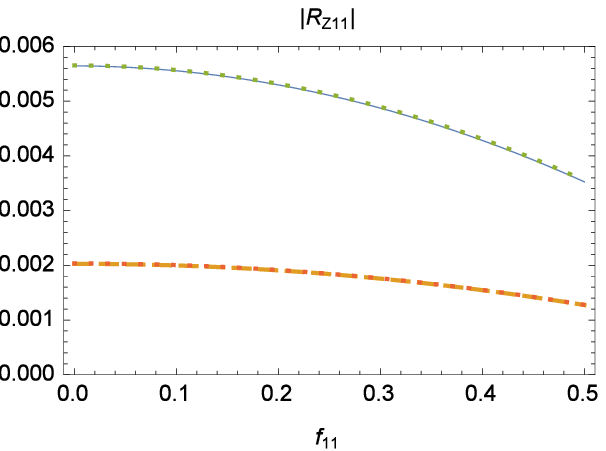}\qquad
\includegraphics[width=6.8cm]{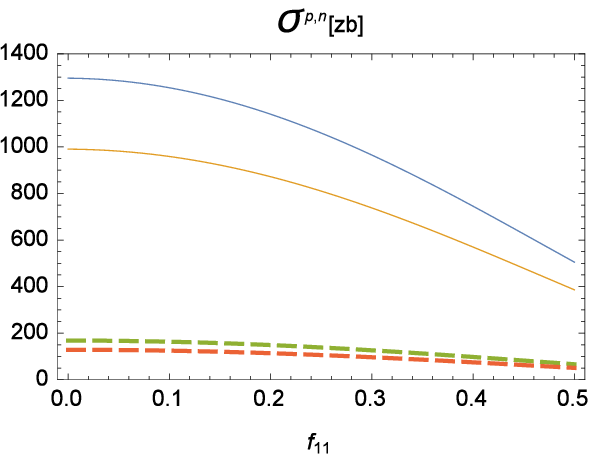}
\caption{({\bf Left}) The~ coupling~ $|R_{Z11}|$~ and~ ({\bf Right}) the cross--section $\sigma^{p,n}$
as a function of $f_{11}$ for $\tilde{f}_{11}=-0.41$, $\tan\beta=2$, $\widetilde{\mu}_1=2\,\mbox{TeV}$,
$\mu_{11} = 200\,\mbox{GeV}$ (solid lines) and $\mu_{11} = 1\,\mbox{TeV}$ (dashed lines).
The upper solid and upper dashed lines represent $\sigma^{p}$ while lower solid and lower dashed lines are associated with $\sigma^{n}$.
The dotted lines correspond to the approximate expression for $R_{Z11}$ (\ref{33}).}
\label{fig20}
\end{figure}

The results of our numerical analysis are presented in Figures~\ref{fig10} and \ref{fig20}.
From these Figures it follows that the approximate expressions (\ref{33}) describe quite well the dependence of the
couplings $|R_{Z11}|$ and $g_{h\chi\chi}$ on the SE$_6$SSM parameters if $|\widetilde{\mu}_1|\gg |\mu_{11}|$.
In particular, $|R_{Z11}|$ and $\sigma^{p,n}$ diminish whereas $g_{h\chi\chi}$ and $\sigma_{SI}$ grow with increasing
$\mu_{11}$ from $200\,\mbox{GeV}$ to $1\,\mbox{TeV}$. The approximate expressions (\ref{33}) indicate that the couplings
$g_{h\chi\chi}$ and $R_{Z11}$ go to zero when $f_{11}\approx - \tilde{f}_{11} \tan\beta$. This means that in the corresponding
part of the parameter space the interactions of $\chi_1$ with the baryons tend to extremely weak.
The vanishing of $g_{h\chi\chi}$ and $R_{Z11}$ can be attained only if the parameters $f_{11}$, $\tilde{f}_{11}$ and $\tan\beta$
are fine-tuned. However as follows from Figures~\ref{fig10} and \ref{fig20} in order to achieve the desirable suppression of
the $\sigma_{SI}$ and $\sigma^{p,n}$ the fine--tuning is not needed. In our analysis
$f_{11}$ is chosen to be positive while $\tilde{f}_{11}$ is fixed to be negative. As a consequence $g_{h\chi\chi}$,
$|R_{Z11}|$, the spin--independent and spin--dependent cross sections decrease when $f_{11}$ grows and approaches
$-\tilde{f}_{11} \tan\beta$. It is worth to point out that in the part of the parameter space of the SE$_6$SSM
where the couplings $|g_{h\chi\chi}|$ and $|R_{Z11}|$ become considerably smaller than $10^{-3}$ one cannot neglect
the contributions to $\sigma_{SI}$ and $\sigma^{p,n}$ which are induced by the heavy Higgs states and $Z'$ boson.
Moreover one should take into account the quantum corrections to the Lagrangian (\ref{26}) that stem from the
one--loop diagrams involving the electroweak gauge bosons \cite{Hisano:2011cs,Hisano:2012wm}. The inclusion of these quantum
corrections lead to $\sigma_{SI}\sim 0.1\,\mbox{yb}$ even if at the tree level $|g_{h\chi\chi}|\ll 10^{-3}$ \cite{Nagata:2014wma}.

\unskip

Now let us compare the computed values of $\sigma_{SI}$ and $\sigma^{p,n}$ with the corresponding experimental bounds
$(\sigma_{SI})_{\text{exp}}$ and $(\sigma^{p,n})_{\text{exp}}$.
The spin--independent $\chi_1$--nucleon scattering cross sections presented in Figure~\ref{fig10} are always
smaller than LUX--ZEPLIN (LZ) experimental limits, i.e. $(\sigma_{SI})_{\text{exp}}\approx 60\,\mbox{yb} (300\,\mbox{yb})$ for
$m_{\chi_1}\approx 200\,\mbox{GeV} (1\,\mbox{TeV})$ \cite{LUX-ZEPLIN:2022qhg}.
The most stringent experimental bound on the spin--dependent WIMP--proton scattering cross section was obtained by the PICO-60 experiment, i.e.
$(\sigma^{p})_{\text{exp}}\approx 10^5\,\mbox{zb}\, (4\cdot 10^5\,\mbox{zb})$ for $m_{\chi_1}\approx 200\,\mbox{GeV} (1\,\mbox{TeV})$ \cite{PICO:2019vsc}.
The spin--dependent WIMP--neutron scattering cross section is more tightly constrained, i.e.
$(\sigma^{n})_{\text{exp}}\approx 0.9\cdot 10^4 \,\mbox{zb}\, (5\cdot 10^4 \,\mbox{zb})$ for $m_{\chi_1}\approx 200\,\mbox{GeV} (1\,\mbox{TeV})$
\cite{LUX-ZEPLIN:2022qhg}. The values of $\sigma^{p,n}$ shown in Figure~\ref{fig20} are considerably smaller than $(\sigma^{p,n})_{\text{exp}}$.

In the SE$_6$SSM the maximal values of the spin--independent and spin--dependent $\chi_1$--nucleon scattering cross sections are
much larger than $\sigma_{SI}$ and $\sigma^{p,n}$ presented in Figures~\ref{fig10} and \ref{fig20}.
These cross sections attain their maximal possible values for $\mu_{11}\simeq \widetilde{\mu}_1$ and $\tilde{f}_{11} \sim f_{11}\sim 1$.
In this case $\sigma_{SI}$ can reach $20-30\,\mbox{zb}$ which is considerably larger than the corresponding experimental limit \cite{LUX-ZEPLIN:2022qhg}.
Since for a given mass of the lightest exotic fermion $\sigma_{SI}$ vanishes when $f_{11} = - \tilde{f}_{11} \tan\beta$,
the spin--independent $\chi_1$--nucleon scattering cross section varies from zero to its maximal value for each $m_{\chi_1}$.
In the scenarios under consideration the suppression of $\sigma_{SI}$ and $\sigma^{p,n}$ is caused by the cancellations of different
contributions to $R_{Z11}$ and $g_{h\chi\chi}$ as well as by the large value of $\widetilde{\mu}_1$ that should be associated with the
sparticle mass scale $M_S$. In the near future the experiments LZ \cite{LUX-ZEPLIN:2018poe}, XENONnT \cite{XENON:2020kmp},
DARWIN \cite{DARWIN:2016hyl} and DarkSide-20k \cite{DarkSide-20k:2017zyg} can set even more stringent limits on
$\sigma_{SI}$ and $\sigma^{p,n}$ constraining further the parameter space of the SE$_6$SSM.

\section{Conclusions}

In this article we considered leptogenesis and the interactions of the dark matter with the nucleons
in the $U(1)_N$ extension of the MSSM, in which the single discrete $\tilde{Z}^{H}_2$ symmetry
forbids flavor--changing transitions and the most dangerous baryon and lepton number violating operators.
The low energy matter content of this SUSY model (SE$_6$SSM) includes three fundamental
$27$ representations of $E_6$, an additional pair of $SU(2)_W$ lepton doublets $L_4$ and
$\overline{L}_4$ with opposite $SU(2)_W \times U(1)_Y \times U(1)_N$ quantum numbers,
a pair of the SM singlet superfields $S$ and $\overline{S}$ with opposite $U(1)_N$ charges
as well as four $E_6$ singlet superfields. Thus the SE$_6$SSM contains extra exotic matter
beyond the MSSM. The scalar components of the superfields $S$ and $\overline{S}$ can
develop VEVs along the D-flat direction, so that $\langle S\rangle \simeq \langle \bar{S}\rangle \gg 10\,\mbox{TeV}$,
breaking the $U(1)_N$ symmetry and inducing TeV scale masses of all extra exotic fermions and $Z’$ boson.
The relatively light components of the supermultiplets $L_4$ and $\overline{L}_4$ allow the
lightest exotic colored state to decay before BBN. They also facilitate the unification of gauge couplings.

In the SE$_6$SSM the cold dark matter density is formed by two stable neutral states.
Here we focused on the scenarios in which one of these stable particles is gravitino.
In this case all TeV scale states can decay before BBN only when gravitino mass $m_{3/2}\lesssim 1\,\mbox{GeV}$.
Another stable state tends to be the lightest exotic neutralino $\chi_1$ with mass $m_{\chi_1}\le 1.1\,\mbox{TeV}$.
Because it is a superposition of the neutral fermion components of the $SU(2)_W$ doublets, the lightest
exotic chargino $\chi^{\pm}_1$, the second lightest exotic neutralino $\chi_2$ and $\chi_1$ are nearly
degenerate around $m_{\chi_1}$. Such scenarios result in the phenomenologically acceptable dark matter
density if the reheating temperatures $T_R\lesssim 10^{6-7}\,\mbox{GeV}$. Even for so low $T_R$ the
decays of the lightest right--handed neutrino/sneutrino in the SE$_6$SSM can generate the appropriate
lepton asymmetry due to the presence of $L_4$ and $\overline{L}_4$ in the particle spectrum. This lepton
asymmetry is converted into the observed baryon asymmetry via sphaleron processes. In the scenarios under
consideration there is a part of the SE$_6$SSM parameter space in which the dark matter--nucleon scattering
cross section is substantially smaller than the present experimental limits.

The phenomenological viability of the scenarios under consideration requires $\chi^{\pm}_1$, $\chi_2$ and $\chi_1$
to be lighter than $1.1\,\mbox{TeV}$. Otherwise the annihilation cross section for $\chi_1+\chi_1 \to \mbox{SM particles}$
becomes too small giving rise to the cold dark matter density which is considerably larger than its measured value.
Relatively light charged and neutral fermions have been searched for in different experiments. If the mass of the lightest
exotic chargino $m_{\chi^{\pm}_1}$ and the mass of the second lightest exotic neutralino $m_{\chi_2}$ are too close to $m_{\chi_1}$
the decay products of $\chi^{\pm}_1$ and $\chi_2$ may escape detection. This happens, for example, within natural SUSY,
where the mass splitting between the lightest and second lightest ordinary neutralino states as well as the mass splitting between
the lightest ordinary chargino and the lightest ordinary neutralino are at least a few GeV \cite{Baer:2012up}--\cite{Baer:2012cf}.
The results of the searches for such degenerate states depend on $\Delta = m_{\chi^{\pm}_1}-m_{\chi_1}$ and $\Delta_1 = m_{\chi_2}-m_{\chi_1}$.

In the scenarios under consideration the SE$_6$SSM parameters are chosen so that $\Delta_1 > 200\,\mbox{MeV}$ while
$\Delta$ is larger than $300\,\mbox{MeV}$ \cite{Nagata:2014wma,Cirelli:2005uq}. Therefore $\chi^{\pm}_1$ and $\chi_2$ cannot be long--lived.
At the LHC the lightest exotic chargino and neutralino states can be produced in pairs via off-shell $W$ and $Z$--bosons. Then
$\chi^{\pm}_1$ and $\chi_2$ subsequently decay into hadrons and $\chi_1$. For $\Delta \simeq 4.7\,\mbox{GeV} (2\,\mbox{GeV})$
ATLAS ruled out $\chi^{\pm}_1$ with masses below $193\,\mbox{GeV} (140\,\mbox{GeV})$ \cite{ATLAS:2019lng}. For $\Delta = 1\,\mbox{GeV}$
CMS excluded $\chi^{\pm}_1$ with masses below $112\,\mbox{GeV}$ \cite{CMS:2019san}. The discovery prospects for such exotic chargino and
neutralino states look more promising at future International Linear Collider (for a review see \cite{Baer:2013cma}).

The SE$_6$SSM also predicts the existence of other exotic neutralino and chargino states. Two exotic neutralino states and the
second lightest exotic chargino are formed by the fermion components of the $SU(2)_W$ doublets. These fermions as well as
their superpartners might be either light or heavy depending on the SE$_6$SSM parameters. Due to the $Z_{2}^{E}$ symmetry conservation
in the collider experiments all exotic particles can only be created in pairs. Since the exotic neutralino and chargino as well as their
scalar partners do not couple to quarks/squarks directly at the LHC these states can be produced via the EW interactions. As a consequence
their production cross section remains relatively small even if the corresponding states have masses around $1\,\mbox{TeV}$.
The conservation of $R$--parity and $Z^{E}_2$ symmetry implies that the final state in the decay of the exotic fermions involves at
least one lightest exotic neutralino while the final state in the decay of its scalar partner should contain at least one lightest
exotic neutralino and one gravitino. If both of the produced states decay into on-shell gauge bosons it is expected that they should
result in some enhancements in the rates of
\begin{equation}
p\,p \to Z\,Z + E_T^{\rm miss} + X\,,\quad p\,p \to W\,Z + E_T^{\rm miss} + X\,,\quad p\,p \to W\,W + E_T^{\rm miss} + X\,,
\label{350}
\end{equation}
where $E^{\rm miss}_{T}$ is associated with the lightest exotic fermion (and gravitino) and $X$ should be identified
with jets and/or extra charged leptons that may stem from the decays of intermediate states.

As mentioned before, the components of $L_4$ and $\overline{L}_4$ are expected to be relatively light.
When all other exotic states and sparticles except $\chi^{\pm}_1$, $\chi_2$, $\chi_1$ and gravitino are rather heavy,
the scalar ($\tilde{L}_4$) and fermionic ($L_4$) components of the supermultiplets $L_4$ and $\overline{L}_4$
can be produced in pairs via off--shell $W$ and $Z$--bosons. Their decays always lead to either $\tau$--lepton or electron/muon
as well as missing energy in the final state. In the case of $\tilde{L}_4$ decays the missing energy in the final state can be
associated with only one lightest exotic neutralino while the final state of the $L_4$ decays has to involve at least one lightest
exotic neutralino and one gravitino to ensure the conservation of $R$--parity and $Z^{E}_2$ symmetry. More efficiently $L_4$
and/or $\tilde{L}_4$ can be produced through the decays of the lightest exotic colored states if these states are relatively light
and the corresponding decay channels are kinematically allowed.

Finally it is worth emphasising that the SE$_6$SSM predicts the existence of extra quarks and their scalar superpartners
that carry lepton and baryon numbers simultaneously \cite{Nevzorov:2012hs}. The LHC lower bounds on the scalar
leptoquark masses \cite{ATLAS:2020dsk}--\cite{CMS:2019ybf} are not directly applicable in this case. Indeed,
ordinary scalar leptoquark with electric charge $-1/3$ decays either to the left--handed neutrino $\nu_j$ and down--type quark $d_i$ or
to charged lepton $\ell_j$ and up--type quark $u_i$. The lightest exotic colored state $q_1$ in the SE$_6$SSM, which is a superposition of
either scalar or fermion components of the supermultiplets $\overline{D}_i$ and $D_i$, is odd under $\tilde{Z}^{E}_2$ symmetry.
As a consequence its decays always lead to the missing energy $E^{\rm miss}_{T}$ in the final state
\begin{equation}
q_1 \to u_i (d_i) + \ell_j (\nu_j) + E^{\rm miss}_{T} + X\,.
\label{35}
\end{equation}
The pair production of the lightest exotic colored states
at the LHC may result in the enhancement of the cross sections of $pp\to jj + E^{\rm miss}_{T} + X$ and/or
$pp\to jj \ell_k \bar{\ell}_m + E^{\rm miss}_{T} + X$. The LHC pair production cross section of the lightest exotic quarks changes from
$10\,\mbox{fb}$ to $1\,\mbox{fb}$ if the mass of $q_1$ increases from $1.3\,\mbox{TeV}$ to $1.7\,\mbox{TeV}$ \cite{Kang:2007ib}.
In the case of the lightest exotic squarks the production cross section is an order of magnitude smaller.
The presence of $Z'$ boson and exotic multiplets of matter in the particle spectrum is a very peculiar feature that should permit
to distinguish the SE$_6$SSM from the MSSM and other extensions of the SM.

\section*{Acknowledgements}
R.N. thanks X.~Tata for very valuable comments and remarks.
R.N. acknowledges fruitful discussions with P.~Athron and L.~Wu.

\end{document}